\documentclass[twocolumn]{aastex6}

\usepackage{verbatim}
\usepackage{aas_macros}
\usepackage{hyperref}
\usepackage{natbib}
\usepackage{graphicx}
\usepackage{epsfig}
\usepackage{amsmath}
\usepackage[english]{babel}

\shorttitle{Stellar, gas and dark matter content of barred galaxies}
\shortauthors{Cervantes Sodi}

\begin{document}

\title{Stellar, gas and dark matter content of barred galaxies}

\author{Bernardo Cervantes Sodi \altaffilmark{1}}
\altaffiltext{1}{Instituto de Radioastronom\'ia y Astrof\'isica, Universidad Nacional Aut\'onoma de M\'exico, Campus Morelia, A.P. 3-72, C.P. 58089 Michoac\'an, M\'exico; b.cervantes@crya.unam.mx}

\begin{abstract}
We select a sample of galaxies from the Sloan Digital Sky Survey Data Release 7 (SDSS-DR7)
where galaxies are classified, through visual inspection, as hosting strong bars, weak bars or as unbarred
galaxies, and make use of HI mass and kinematic information from the Arecibo Legacy Fast ALFA
(ALFALFA) survey catalog, to study the stellar, atomic gas and dark
matter content of barred disk galaxies. 
We find, in agreement with previous studies, that the bar fraction increases
with increasing stellar mass. A similar trend is found with total baryonic mass, although 
the dependence is not as strong as with stellar mass, this due to the contribution of gas.
The bar fraction shows a decrease with increasing gas mass fraction. This anticorrelation between
the likelihood of a galaxy hosting a bar with the gas richness of the galaxy results from the
inhibiting effect the gas has in the formation of bars. We also find that for massive galaxies with stellar masses
larger than 10$^{10} M_{\odot}$, at fixed stellar mass, the bar fraction decreases with
increasing global halo mass (i.e. halo mass measured up to a radius of the order of the HI disk extent).

\end{abstract}

\keywords{
galaxies: fundamental parameters --- galaxies: halos --- galaxies: spiral
--- galaxies: statistics --- galaxies: structure}

\section{Introduction}

In the local Universe, a substantial percentage of massive
galaxies are known to present stellar bars (e.g. de Vaucouleurs et al. 1991; 
Eskridge et al. 2000; Laurikainen, Salo \& Buta 2004; Buta et al. 2010;
Nair \& Abraham 2010; Buta et al. 2015; Cervantes Sodi et al. 2015, henceforth CS+15; Gavazzi et al. 2015).

Given the non-axisymmetric nature of bars, they are expected to speedup
secular evolution in galaxies, leading to mass and angular momentum
redistribution within the components of the galaxies (Lynden-Bell 1979; Roberts, Huntley \& van Albada
1979; Sellwood 1981; Weinberg 1985; Sellwood \& Wilkinson 1993;
Athanassoula 2003; Kormendy \& Kennicutt 2004; Cheung et al. 2013; Sellwood 2014).

The origin of stellar bars has been addressed from the first numerical simulations,
starting with the pioneering work by Ostriker \& Peebles (1973) where they
simulated galaxies with hundreds of mass points, and found that all of their
simulated systems were unestable to barlike modes, but the inclusion of a
spherical halo component with a halo-to-disk mass ratio larger than 1, was enough
to prevent the formation of bars. They proposed a stability criterion based
on the ratio of the kinetic energy of rotation over total gravitational energy, with
a marginal value of 0.14 for reaching stability.
In a similar way, Efstathiou, Lake \& Negroponte (1982), through numerical experiments,
proposed their own stability criterion in terms of the maximum rotation curve velocity ($\textit{v}_\mathrm{max}$),
and the mass and scale-length of the disk ($M_\mathrm{d}$, $f_\mathrm{d}$), defined as
$\epsilon_c=v_\mathrm{max}/(GM_\mathrm{d}/r_\mathrm{d})^{1/2} > 1.1$ for stable systems.
In this sense, $\epsilon_c$ gives a measure of the self-gravity of the disk
(see also Christodoulou et al. 1995).
Athanassoula \& Sellwood (1986), using 2D simulations, confirmed the analytical result
by Toomre (1981), that a higher halo-to-disk mass ratio decreases the bar growth
rate, but also pointed out the relevance of random motions within the disk, that
also help to stabilize the disk against bar formation, even for maximal disks.

The effects of the halo on bar formation and growth are not as simple as they seemed
in early works. For instance, the strength of the bar and the decrease of its pattern speed
is set by the amount of angular momentum that is able to loose. A responsive dark matter halo can
work as a sink of angular momentum, allowing the growth of bars in the secular evolution phase, although
in the formation phase the presence of a massive halo slows down their formation
(Athanassoula 2002; 2003 and Athanassoula 2013 for a review).
Debattista \& Sellwood (2000) concluded that in order to maintain their observed high pattern speeds,
bars must be hosted by low central density halos, because for dens halos the drag force due to dynamical friction
between the bar and the halo is enough to drive the corotation point out to unrealistic distances, in
this way, not only the mass of the halo is relevant, but also its density.
The triaxiality of halos produce significant effects on the origin and fate of bars. Triaxial halos
induce early bar formation (Berentzen et al. 2006; Athanassoula et al. 2013), but once the bar is formed, they
damp their growth (Berentzen et al. 2006; Machado \& Athanassoula 2010; Athanassoula et al. 2013). If a rotating halo is
included, the growth in size and strength of bars gets quenched with increasing spin
(Saha \& Naab 2013; Long et al. 2014), which explains why the bar fraction decreases
with increasing spin (Cervantes Sodi et al. 2013).

Although the interaction between the halo component and the bar is complex,
results from recent simulations coincide on that the disk-to-halo mass ratio
is a factor of primary importance in bar formation and evolution. In some cases, bar
formation is suppressed if the halo mass is increased (DeBuhr et al. 2012; Yurin \& Springel 2015), while in
some other cases, the bars are produced even in simulated galaxies with
low disk-to-halo mass ratios, but the amplitude of the bars is smaller in halo
dominated systems, and the growth of the bar slows down (Sellwood 2016).

The effects of the disk-to-halo mass ratio on bar formation and evolution, have also been
studied from an observational perspective. Working with a sample of bright barred and
unbarred galaxies, Courteau et al. (2003) found that for a given luminosity, the structural
and dynamical parameters of the two subsamples are comparable, with barred and
unbarred galaxies following the same Tully-Fisher relation, which implies that at fixed
luminosity, barred and unbarred galaxies have halos of comparable mass. More recently,
using a volume-limited sample of galaxies from the SDSS, with halo masses taken
from the Yang et al. (2007) galaxy group catalog, CS+15 found a strong correlation
between the fraction of galaxies hosting strong bars and the stellar-to-halo mass ratio,
with f$_\mathrm{bar}$ increasing with increasing $M_\mathrm{*}/M_\mathrm{halo}$,
even at fixed stellar mass. D\'iaz-Garc\'ia et al. (2015) found a similar dependence of
the bar fraction with $M_\mathrm{*}/M_\mathrm{halo}$, but the dependence in
their sample vanishes at fixed stellar mass. This question will be addressed
in the present work.

Observational studies have shown that bars are more frequently found in
massive, red galaxies with early-type morphologies and prominent bulges (Erwin 2005; Sheth et al.
2008; Weinzirl et al. 2009; Nair \& Abraham 2010; Hoyle et al. 2011; Lee et al 2012a, henceforth Lee+12;
Cervantes Sodi et al. 2013), highlighting that the formation of bars is strongly
dependent on the physical properties of the hosting galaxy. 
Results by Sheth et al. (2008) and Kraljic, Bournaud \& Martig (2012) show a dramatic decline
of the bar fraction with increasing redshift for low-mass galaxies, while the bar fraction in massive, luminous
galaxies remains constant out to $z \sim 0.8$. This suggests that bars
form later in low mass galaxies, in a downsizing way, with bars forming at the same epoch at
which galaxies become kinematically cold, dominated by a thin stellar disk.
One crucial component
in the formation process and evolution of bars is the fraction of mass in form of gas in the galaxy.
When a galaxy is rich in gas, a significant exchange  of angular momentum
is expected to occur between the stellar and gas components (Friedli \& Benz 1993; Athanassoula 2003;
Combes 2008). In this exchange, the angular momentum lost by the gas
is transferred to the bar, letting the gas inside the co-rotation radius to fall to the central region and
preventing the inflow of gas from external regions (Athanassoula 1992; Heller \& Shlosman 1994;
Knapen et al. 1995; Sheth et al. 2005).
At the same time, the interchange of angular momentum increases the rotation frequency of the bar,
weakening it and ultimately destroying it (Combes 2008).
Recent simulations including feedback, cooling and star formation, (Athanassoula, 
Machado \& Rodionov 2013) have also shown that bars in the presence of large amounts
of gas are expected to form later, and at all times are weaker, than in gas-poor simulations.

The effect of the gas on the structure of the bar can be due to an indirect process,
as suggested by Berentzen et al. (2007), with the bar fuelling gas to the center of the
galaxy, where it grows a central mass concentration that weakens the bar.
This inflow of gas to the center can in turn produce a central starburst as
predicted by simulations (Shlosman, Frank \& Begelman 1989; Berentzen et al. 1998)
and confirmed by some observational studies (Laurikainen, Salo \& Buta 2004;
Jogee, Scoville \& Kenney 2005; Ellison et al. 2011; Wang et al. 2012).
At the end, the result of enhanced central star formation in barred galaxies,
and the adverse effects of the gas on the evolution of the bar, could
explain why the bar fraction observed in local galaxies
decreases for increasing gas mass fraction (Masters et al. 2012, Cheung et al. 2013).

In this paper we study the dependence of the bar fraction on the stellar-to-halo mass
ratio following CS+15 but using a more direct approach to estimate halo masses. Instead
of using a method that depends on the clustering of galaxies and assumes a one-to-one relation
between the total luminosity of the groups and the halo mass, here we explore a more direct approach,
assigning halo masses using kinematic information from HI line widths.
We also study the dependence of the bar fraction on the gas mass ratio with the aim to
disentangle if the decrease of the bar fraction on gas rich galaxies is caused by
bars promoting the consumption of gas and/or, if the increase of gas content
inhibits bar formation. The main results and discussion
are presented in Section 3. Lastly, we summarize our general conclusions in Section 4.
Throughout this paper, we use a cosmology with density parameter
$\Omega_{\mathrm{m}}$ = 0.3, cosmological constant
$\Omega_{\mathrm{\Lambda}}$ = 0.7 and Hubble constant written as
$H_{0}  = 100h$km s$^{-1}$ Mpc$^{-1}$, with $h=0.7$.

\section{Data}

\subsection{Galaxy Catalog}

\begin{figure*}
\label{distributions}
\centering
\begin{tabular}{cc}
\includegraphics[width=0.4\textwidth]{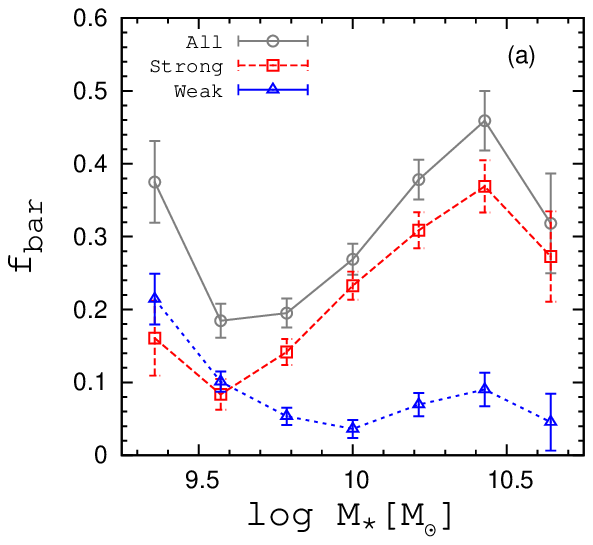} & \includegraphics[width=0.4\textwidth]{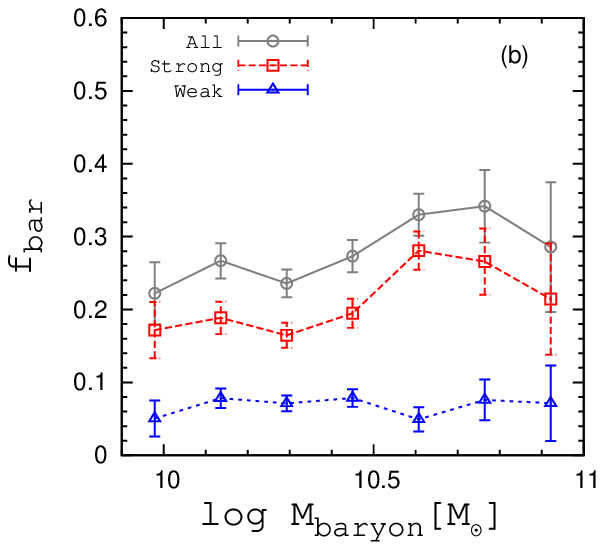} \\
\includegraphics[width=0.4\textwidth]{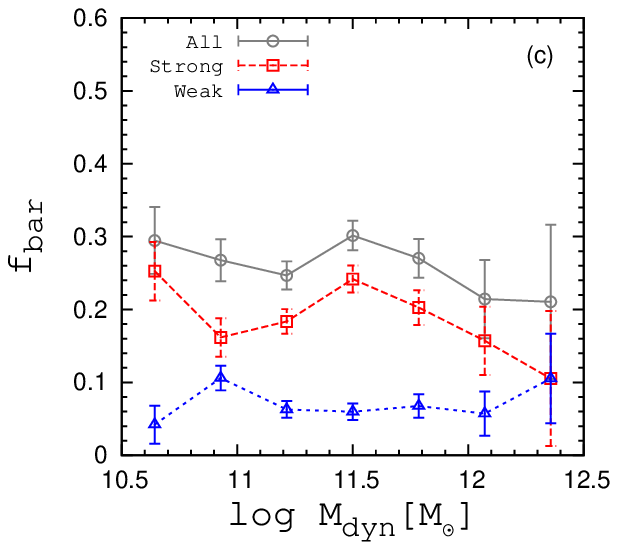} & \includegraphics[width=0.4\textwidth]{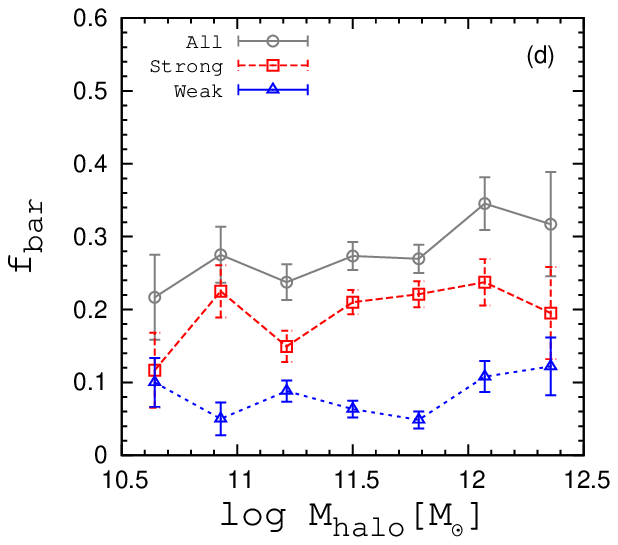}
\end{tabular}
\caption{The fraction of barred galaxies $f_{\mathrm{bar}}$ as a function of:  (\textit{a})
stellar mass M$_{\mathrm{*}}$, (\textit{b}) baryonic mass M$_{\mathrm{baryon}}$,  (\textit{c})
dynamic mass M$_\mathrm{dyn}$, and (\textit{d}) halo mass M$_\mathrm{halo}$.
}\label{Masses}
\end{figure*}

The parent galaxy sample used in this work consist of nearly $\sim$30,000 galaxies
(Lee+12) drawn from the Sloan Digital Sky Survey Data Release 7 (DR7; Abazajian et al. 2009).
It is selected as a volume-limited sample, within the redshift
range 0.02 $\leq z \leq$ 0.05489, and complete down to a limit
$r$-band absolute magnitude brighter than $M_{\mathrm{r}} =$  -19.5 + 5log$\mathrm{h}$.
The morphological classification of the sample is obtained using the
method developed by Park \& Choi (2005), where galaxies are segregated
into early- and late-types in the color vs. color gradient and concentration index
planes. For the present study we keep only late-type galaxies. 

The bar identification is obtained by visual inspection of combined
color images of three SDSS bands ($g+r+i$), and once a bar is
identified is further classified as a strong bar if its length is
larger than one quarter of the optical size of the host galaxy,
or as a weak bar otherwise. We further
restrict the sample to those galaxies with
$i$-band isophotal axis ratio $b/a > 0.6$, $a$ and $b$ being the 
semi-major and semi-minor axes, this in order to avoid selection
biases by inclination, given that bars are easier to identify in face-on
galaxies. Keeping only late-type galaxies mostly face-on reduces
the parent sample to 10,674 galaxies, of which 23.8\% host
strong bars and 6.5\% host weak bars, giving a total bar fraction of
30.4\%. A detail comparison of this volume limited sample is presented in
section 3.2 of Lee+12, where the authors show a
 good agreement with the classification performed by Nair \& Abraham (2010).
 Furthermore, the dependence of the bar fraction of the sample on
 stellar mass, color and concentration index are in qualitative and quantitative
 good agreement with the findings by Nair \& Abraham (2010),
 Masters et al. (2011) and Oh et al. (2012).
We refer the reader to Lee+12
for a more detailed description of the sample, along with
a number of studies that make use of this same sample (Lee et al. 2012b;
Cervantes Sodi et al. 2013; Lin et al. 2014; Cervantes Sodi et al. 2015). 

The optical photometric properties required for this study are
extracted from the Korea Institute for Advanced Study Value-Added
Galaxy Catalog (Choi et al. 2010), together with the NASA Sloan
Atlas3 (NSA) catalog (Blanton et al. 2011), from where we get the stellar
masses of our galaxies. The stellar masses in the NSA are calculated
using SDSS and GALEX photometric bands and the $kcorrect$ software
by Blanton \& Roweis (2007), assuming a Chabrier (2003) initial mass function.
The estimations of star formation rates (SFR) and specific star formation rates (sSFR)
come from the the New York University Value Added Galatic Catalog (NYU-VAGC; Blanton et al. 2005)
and the MPA/JHU SDSS database (Kauffmann et al. 2003; Brinchmann et al. 2004).

\subsection{HI data}

\begin{figure*}
\label{distributions}
\centering
\begin{tabular}{cc}
\includegraphics[width=0.4\textwidth]{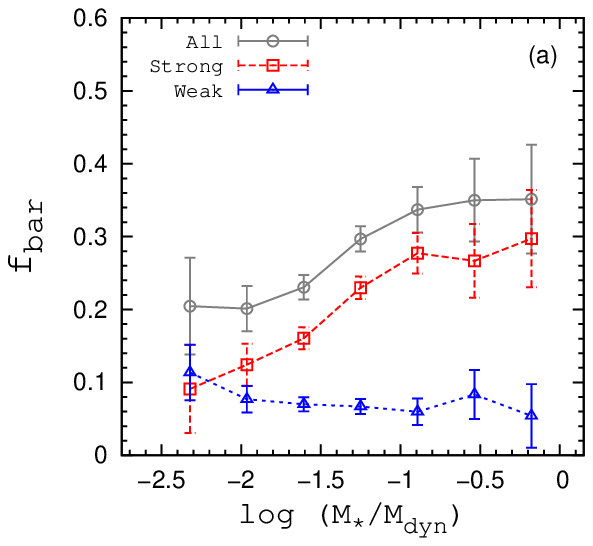} & \includegraphics[width=0.4\textwidth]{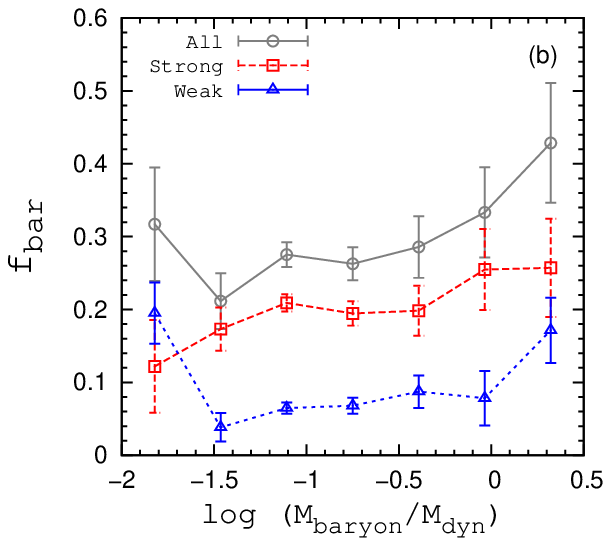} \\
\includegraphics[width=0.4\textwidth]{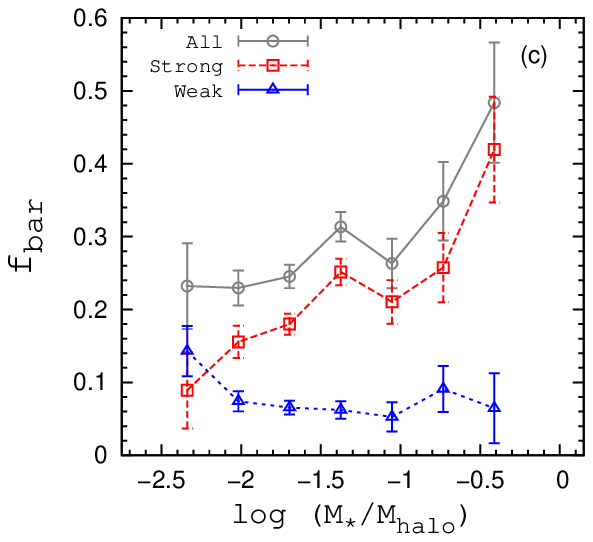} & \includegraphics[width=0.4\textwidth]{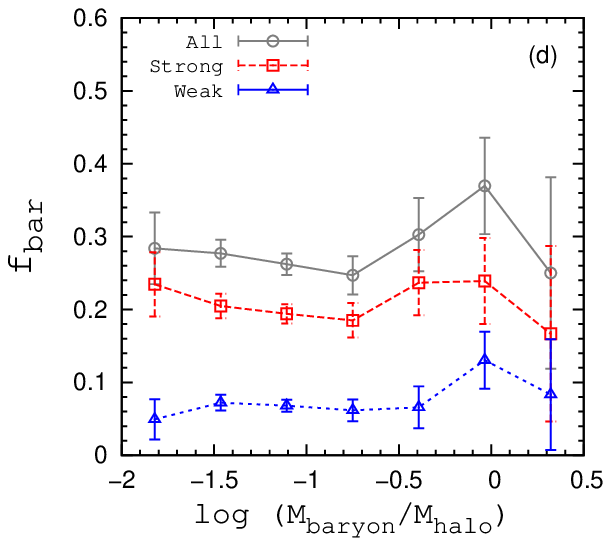}
\end{tabular}
\caption{The fraction of barred galaxies $f_{\mathrm{bar}}$ as a function of:  (\textit{a})
stellar-to-dynamic mass ratio M$_{\mathrm{*}}/$M$_{\mathrm{dyn}}$, (\textit{b}) baryonic-to-dynamic mass ratio M$_{\mathrm{baryon}}/$M$_{\mathrm{dyn}}$,  (\textit{c})
stellar-to-halo mass ratio M$_{\mathrm{*}}/$M$_{\mathrm{halo}}$, and (\textit{d}) bayonic-to-halo mass ratio M$_{\mathrm{baryon}}/$M$_{\mathrm{halo}}$.
}\label{Fractions}
\end{figure*}

To estimate dynamical masses and halo masses for the galaxies in our sample,
as well as gas mass fractions, we turn to the ALFALFA survey (Giovanelli et al. 2005), which is a blind,
single-dish, flux-limited extragalactic HI survey. We use data from the publicly available
catalog of the ALFALFA survey "$\alpha$.70", which covers about 70$\%$ of the final survey area,
which is planed to be 7,000 deg$^2$ of high galactic latitude sky observable with the telescope
and an expected detection of $>$ 30,000 galaxies, out to $cz \approx$ 18,000 km s$^{-1}$.
A description of the "$\alpha$.40" catalog can be found in Haynes et al. (2011).

HI masses are estimated using the formula by Haynes \& Giovanelli 1984;

\begin{equation}
M_\mathrm{HI}=2.356 \times 10^{5} \left( \frac{D}{\mathrm{Mpc}} \right)^{2} \frac{S_\mathrm{21}}{\mathrm{Jy \  km \ s^{-1}}} [M_{\odot}]
\end{equation}

where $S_\mathrm{21}$ is the integrated HI line flux density in units of Jy km s$^\mathrm{-1}$, and $D$ is the distance
in Mpc. From the HI mass, we estimate the atomic gas mass as $M_\mathrm{gas} = 1.4M_\mathrm{HI}$, where the
numeric factor $1.4$ is introduced to account for the presence of helium.
The baryonic mass is simply $M_\mathrm{baryon} = M_\mathrm{gas} + M_\mathrm{*}$.

To calculate dynamical masses for the galaxies in the sample, we first estimate the galactic
rotational velocity as

\begin{equation}
V_\mathrm{rot}=W/(2 \times sin \ i),
\end{equation}

where $W$ is the HI line width measured at 50\% of the peak flux, in units of km s$^\mathrm{-1}$, as
provided by the ALFALFA $\alpha$.70 catalog, and $i$ is the inclination angle of the galaxy, which
is computed through the expression:

\begin{equation}
cos^2 i = \frac{(b/a)^2-q_{0}^2}{1-q_{0}^2},
\end{equation}

with $q_{0}=0.13$ assumed for the intrinsic axial ratio of galaxies viewed edge-on (Giovanelli et al. 1994).
The output of the match of our parent sample with the $\alpha$.70 catalog gives a total of 1,471 galaxies for
our study, which is primarily limited by the ALFALFA detection limits,
imposing a bias towards gas-rich systems. From the total sample, 293 (20\%) are galaxies
hosting strong bars, 106 (7\%) host weak bars and 1,072 (73\%) are unbarred galaxies. The low fraction
of barred galaxies in the sample is due to the strict classification criteria used to identify bars and the bias towards
gas-rich galaxies, that as we show in section 3.3, present a lower bar fraction when compared with
gas-poor systems.

\section{Results and discussion}

\subsection{Bar fraction as a function of mass}

We start our analysis by looking at the dependence of the bar fraction $f_\mathrm{bar}$ as a function
of different masses defined for our sample. In Figure~\ref{Masses}a, we show the dependence of
the bar fraction on stellar mass for strong and weak bars, as well as for strong plus weak bars, 
with the well known trend of higher $f_\mathrm{bar}$ for galaxies
with high stellar masses (Masters et al. 2012; Oh et al. 2012; Skibba et al. 2012; Cervantes Sodi et al. 2013;
Gavazzi et al. 2015), at least for the case of strong bars, an expected result given that
bars form earlier in massive galaxies, as previously shown by Sheth et al. (2008) and Kraljic, Bournaud \&
Martig (2012). Error bars in all figures
denote the estimated 1$\sigma$ confidence intervals based on the bootstrapping resampling
method.
The corresponding result using the baryonic mass is presented in Figure~\ref{Masses}b,
where is noticeable a slight increase of the bar fraction for increasing baryonic mass, but
the trend is less dramatic than the one present as a function of stellar mass.
This might be
due to the fact that an increase in $M_\mathrm{baryon}$ can be the result of an increase of $M_\mathrm{*}$
but also and increase of $M_\mathrm{gas}$, and as will be discussed in section 3.3, an increase
of $M_\mathrm{*}$ promotes the growth of the bar, but an increase of $M_\mathrm{gas}$
hinders the growth of the bar.

To study the dependence of the bar fraction on the halo mass, following Bradford, Geha \& Blanton
(2015), we employ two different estimates.
We calculate the dynamical mass $M_\mathrm{dyn}$, as the mass responsible for establishing a
flat rotation curve with amplitude $V_\mathrm{rot}$ within the HI disk radius;

\begin{equation}
M_\mathrm{dyn}=\frac{R_\mathrm{HI}V_\mathrm{rot}^2}{G}.
\end{equation}

Given that we do not count with rotation curves, we follow Broeils \& Rhee (1997) 
to estimate the radius of the HI disk ($R_\mathrm{HI}$) in terms of the HI mass
using one of the tightest scaling relations of galaxy disks (Lelli, McGaugh \& Schombert 2016):

\begin{equation}
log \ M_\mathrm{HI} = 1.96 \ log \ D_\mathrm{HI} + 6.52,
\end{equation}

with $R_\mathrm{HI}=D_\mathrm{HI}/2$. By using equation 4, we are measuring the halo mass
to a distance from the center that extends to the radius of the HI disk, which typically extends to
twice the optical size of the galaxies (Broeils \& Rhee 1997).
The fraction of barred galaxies as a function of dynamical mass is shown in Figure \ref{Masses}c,
where $f_\mathrm{bar}$ seems to be independent of $M_\mathrm{dyn}$.

As a second estimate for the halo mass we turn to the study by van den Bosch (2002), where he
explored different virial mass estimators for disk galaxies using models for the formation of these
kind of galaxies. The best estimator, with the smallest scatter, is a combination of circular velocity
and disk scale radius $r_\mathrm{d}$, of the form:

\begin{equation}
M_\mathrm{halo}=2.54 \times 10^{10} M_\odot \left( \frac{r_\mathrm{d}}{\mathrm{kpc}} \right) \left( \frac{V_\mathrm{rot}}{100 \  \mathrm{km \ s^{-1}}} \right)^{2}.
\end{equation}

The result using $M_\mathrm{halo}$ (Figure \ref{Masses}d) is very similar to the one using
$M_\mathrm{dyn}$, with little or no dependence of the bar fraction on either of these
mass estimates, a result in agreement with Mart\'inez \& Muriel (2009) and Wilman \& Erwin (2012)
who found no evidence of bars preferring any particular halo mass.

\subsection{Bar fraction as a function of stellar and baryonic fractions}

\begin{figure}
\label{distributions}
\centering
\begin{tabular}{cc}
\includegraphics[width=0.45\textwidth]{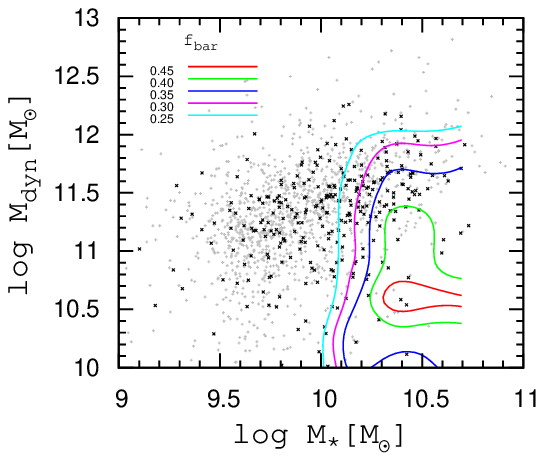}  \\
\includegraphics[width=0.45\textwidth]{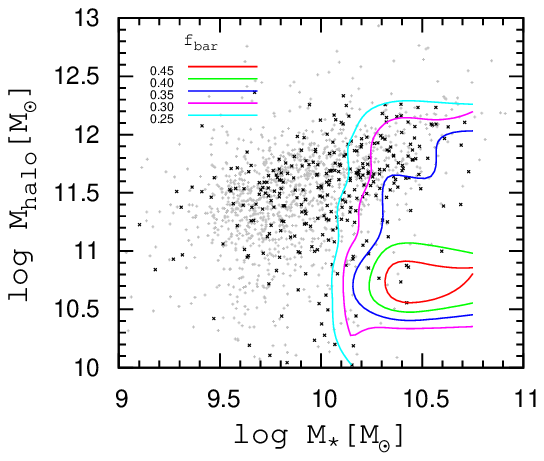} 
\end{tabular}
\caption{Bar fraction $f_\mathrm{bar}$ isocontours in the M$_\mathrm{dyn}$ vs. M$_\mathrm{*}$ (top panel)
and M$_\mathrm{halo}$ vs. M$_\mathrm{*}$ (bottom panel) for strong bars. The contours denote regions of
constant $f_\mathrm{bar}$ in the range $0.25 \leq f_\mathrm{bar} \leq 0.45$. Gray dots represent unbarred
galaxies, black dots represent barred ones.
}\label{Contours}
\end{figure}

In CS+15, the authors showed that a strong dependence is found for
the bar fraction with the stellar-to-halo mass fraction, with the dependence present for strong bars in massive
systems even at fixed stellar mass. In this section we explore if our sample presents this same
dependence. Figures \ref{Fractions}a,c present the bar fraction as a function of the
 stellar-to-dynamic mass and the stellar-to-halo mass ratio respectively, showing the same
 behavior as the one reported by CS+15 by a totally independent method, with the fraction of strong bars
 increasing systematically with increasing the mass ratio. 
 
 Having estimated the baryonic mass for the galaxies of the sample, we also present the bar fraction
 as a function of baryonic-to-dynamic mass and baryonic-to-halo mass ratios in Figure \ref{Fractions}b,d.
 Figure \ref{Fractions}b shows a weak dependence of the bar fraction for strong bars on the baryonic-to-dynamic
 mass ratio, much weaker than the one presented in Figure \ref{Fractions}a with the stellar-to-dynamic mass ratio.
 Finally, Figure \ref{Fractions}d shows no dependence of the bar fraction on the baryonic-to-halo mass ratio.
 The difference in the behaviour of $f_\mathrm{bar}$ with the mass ratios using stellar mass and baryonic
 mass comes from the inclusion of the gas component, which we will explore in more detail in \ref{GasSection}.
 
 The increase of $f_\mathrm{bar}$ with increasing $M_\mathrm{*}/M_\mathrm{halo}$
and $M_\mathrm{*}/M_\mathrm{dyn}$, given the little dependence of the bar fraction
on $M_\mathrm{dyn}$ and $M_\mathrm{halo}$, must be preferentially driven by the dependence
on $M_\mathrm{*}$. In what follows, we will explore the dependence of the bar fraction on the halo mass
at fixed stellar mass.
 
In Figure 2a of CS+15, the authors show that the bar
fraction in the M$_\mathrm{halo}$ vs. M$_\mathrm{*}$ plane presents a dependence on the
halo mass even at fixed stellar mass, specially for the case of strong bars where the effect
is clear, with $f_\mathrm{bar}$ increasing for decreasing M$_\mathrm{halo}$ at fixed M$_\mathrm{*}$.
For the case of weak bars, the dependence is weak and goes in the opposite direction.

D\'iaz-Garc\'ia et al. (2016) also explored the dependence of the bar fraction on the
stellar-to-halo mass ratio using a smaller sample than the one used by CS+15, but
their estimate of $M_\mathrm{*}/M_\mathrm{halo}$ is a more direct one that the one used
by CS+15, comparing the circular
velocity curve infered from infrared images with the
inclination-corrected HI velocity amplitude as obtained from various sources (Courtois et al. 2009; 2011
and HyperLEDA). When using Fourier and ellipse fitting methods to detect bars,
D\'iaz-Garc\'ia et al. (2016) obtain the same tendency of $f_\mathrm{bar}$ decreasing with increasing $M_\mathrm{halo}/M_\mathrm{*}$
ratio, but they report that this tendency is suppressed at fixed stellar mass.

To test if in our sample the dependence of $f_\mathrm{bar}$ on $M_\mathrm{*}/M_\mathrm{halo}$
and $M_\mathrm{*}/M_\mathrm{dyn}$ vanishes at fixed stellar mass, we follow CS+15 exploring
the behavior of the strong bar fraction in the M$_\mathrm{halo}$ vs. M$_\mathrm{*}$ and
M$_\mathrm{dyn}$ vs. M$_\mathrm{*}$ planes. To get a smooth transition of $f_\mathrm{bar}$,
we use a spline kernel, dividing the plane in 10 $\times$ 10 regions where $f_\mathrm{bar}$ is
estimated, requiring a minimum of 10 galaxies per region. This combination is chosen after testing
different bin sizes and minimum of galaxies requested on each bin, by generating mock samples from the
same parent sample and varying the combination of these two parameters until the contours cease to fluctuate from
sample to sample. The contours in Figure \ref{Contours}
show that the maximum value of $f_\mathrm{bar}$ is reached in the region denoted by  M$_\mathrm{*} > 10^{10.25}$ M$_{\odot}$
and M$_\mathrm{dyn}$, M$_\mathrm{halo} < 10^{11}$ M$_{\odot}$. The contours in both panels of Figure \ref{Contours}
also show that there is an increase of $f_\mathrm{bar}$ with increasing M$_\mathrm{*}$,
and for massive galaxies with M$_\mathrm{*}>10^{10}$ M$_{\odot}$, an increase of the bar fraction with
decreasing M$_\mathrm{dyn}$ and M$_\mathrm{halo}$, showing that even at fixed stellar mass,
there is a dependence of the bar fraction on M$_\mathrm{dyn}$ and M$_\mathrm{halo}$.

\begin{figure}

\includegraphics[width=.475\textwidth]{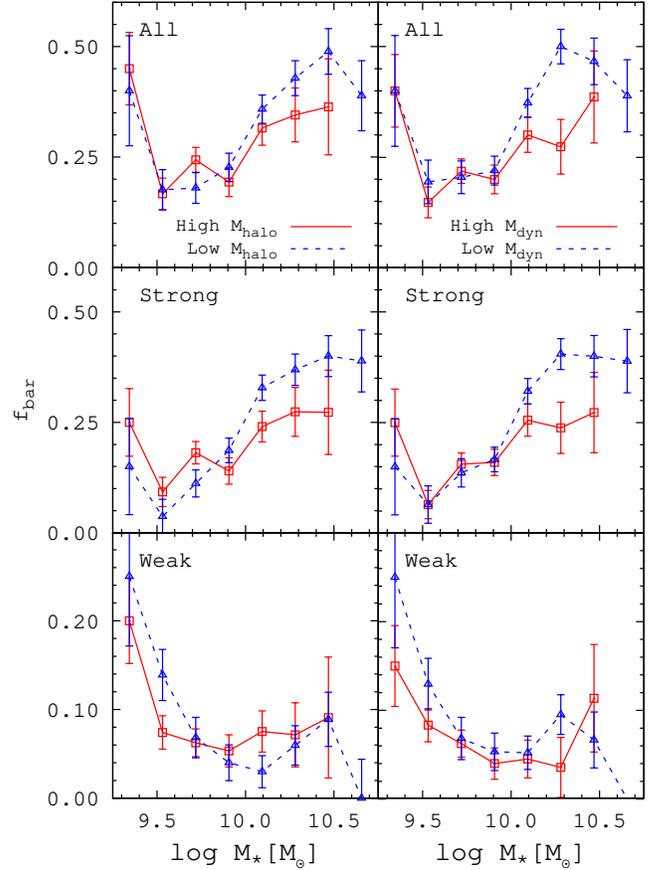} 

\caption{The fraction of barred galaxies $f_{\mathrm{bar}}$ as a function of
M$_{\mathrm{*}}$ for the full sample (strong plus weak bars, top panels), strong bars
(middle panels) and weak bars (bottom panels). For each bin a pair of values is
plotted, in solid (red) lines galaxies with high M$_{\mathrm{halo}}$,
values and in broken (blue) lines galaxies with low M$_{\mathrm{halo}}$
values. Right panels present the same as left panels but the sample is segregated using
M$_{\mathrm{dyn}}$ instead of M$_{\mathrm{halo}}$}
\label{Stellar_div}
\end{figure}

An other way to present this same result is looking at the bar fraction as a function
of stellar mass dividing the sample in two according to its dynamic mass and halo mass.
In Figure \ref{Stellar_div} left panels is shown the bar fraction as a function of stellar mass for the 
full sample in the top panel, strong bars in the middle panel and weak bars in the bottom panel.
Each bin is further divided in half according to their dynamic mass, the solid (red) line corresponds to 
the subsample with high M$_\mathrm{halo}$ values while the broken (blue) to the subsample with
low M$_\mathrm{dyn}$ values. The top and middle panels show that massive
galaxies (M$_\mathrm{*}\geq10^{10}$ M$_{\odot}$) with low M$_\mathrm{halo}$
values have systematically a higher bar fraction
than galaxies with high M$_\mathrm{halo}$.
For the case of weak bars, no systematic difference is found between the two subsamples.

The right panels of Figure \ref{Stellar_div} present the same analysis but using M$_\mathrm{dyn}$
instead of M$_\mathrm{halo}$, showing the same systematic variation but with greater
significance, in good agreement with the results by CS+15. This result also goes in line with recent
simulations  where is found that the disk-to-halo mass
ratio is a factor of primary importance in bar formation and evolution. In some of these studies
\citep{DeBhur12,Yurin15}, bar formation is suppressed if the halo mass is increased, while
in other cases bars are formed even in simulated galaxies with low disk-to-halo mass
ratios, but the amplitude of the bars is smaller in halo dominated systems \citep{Sellwood16}.

We recall that we are dealing with global, integrated properties
of galaxies, and that this strong dependence of the bar fraction with the stellar-to-halo mass
and the stellar-to-dynamic mass ratio might change if we focus on the mass contribution
of the halo in the central regions of the galaxy, given that it is only the mass in the relatively inner
part of the halo that can interact with the bar (Athanassoula \& Misiriotis 2002). In this regard,
the results by Cervantes Sodi et al. (2013) are complementary. They show that the bar
fraction decreases for increasing spin, and if low surface brightness galaxies reside in halos
with large values of spin (Jimenez et al. 1998; Kim \& Lee 2013), 
being these galaxies dark matter dominated even in the central
regions, then galaxies dominated by their dark matter component in the inner parts
appear to be more stable against bar formation. The results shown in the present study in
combination with the results by CS+15 and Cervantes Sodi et al. (2013), suggest that gravitational dominant
stellar disks, both, in the inner parts of the galaxy as well as globally, are more prone to
develop and maintain stellar bars, in good agreement with what reported by Algorry et al.
(2016) studying the formation of 269 disks in a $\Lambda$CDM cosmological hydrodynamical
simulation from the EAGLE project, where 82$\%$ of their identified strong bars are hosted
by simulated galaxies that satisfy a combined criteria of presenting a gravitational dominant disk
within the half-mass radius, and have a declining rotation curve
beyond the outer confines of the disk.

 \subsection{Bar fraction as a function of HI richness} \label{GasSection}

\begin{figure}
\label{distributions}
\centering
\begin{tabular}{cc}
\includegraphics[width=0.4\textwidth]{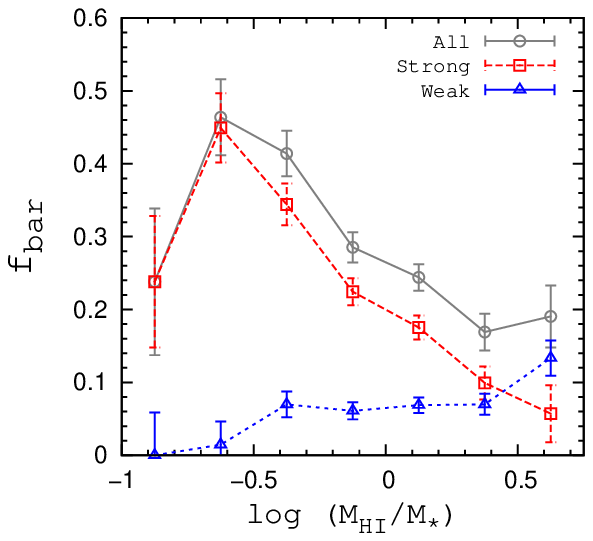} \\
\includegraphics[width=0.45\textwidth]{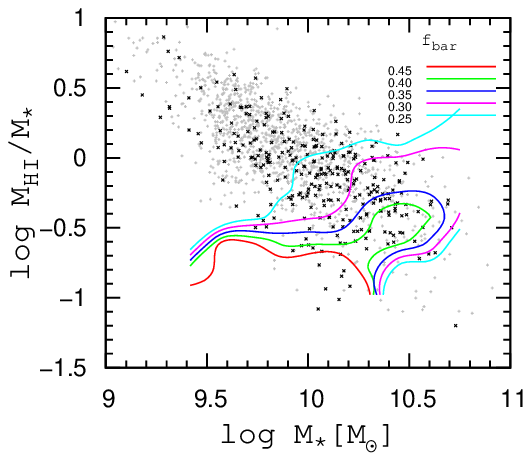} 
\end{tabular}
\caption{(Top panel) Bar fraction as a function of HI gas mass fraction for strong, weak and
strong plus weak bars in our sample. (Bottom panel) Strong bar fraction in the $M_\mathrm{HI}/M_\mathrm{*}$ vs.
$M_\mathrm{*}$ plane. The contours denote regions of
constant $f_\mathrm{bar}$ in the range $0.25 \leq f_\mathrm{bar} \leq 0.45$. Gray dots represent unbarred
galaxies, black dots represent barred ones.
}\label{fgas}
\end{figure}

\begin{figure}
\label{distributions}
\centering
\begin{tabular}{cc}
\includegraphics[width=0.4\textwidth]{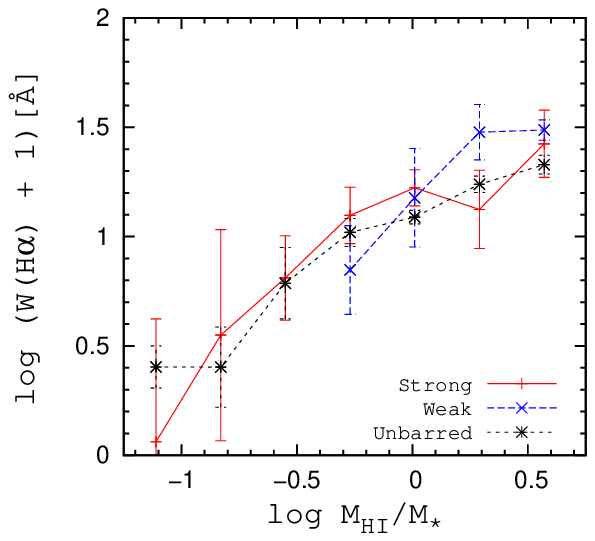}  \\
\includegraphics[width=0.4\textwidth]{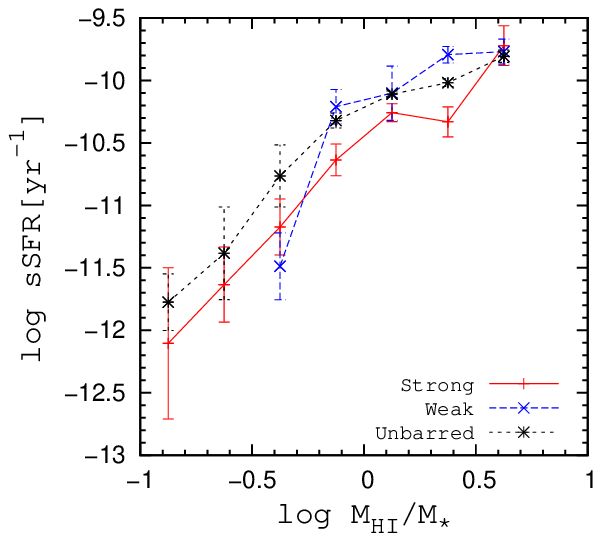}  \\
\includegraphics[width=0.4\textwidth]{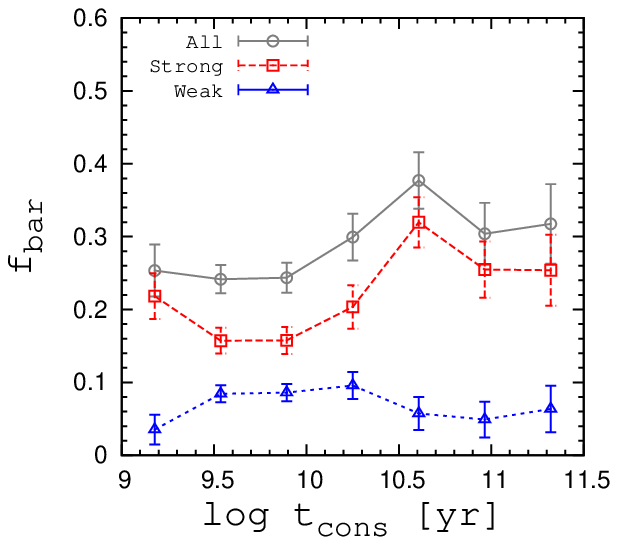} 
\end{tabular}
\caption{(Top panel) Dependence of the equivalent width of H$\alpha$ line on $M_\mathrm{HI}/M_\mathrm{*}$
for galaxies with strong bars, weak bars and unbarred galaxies. (Middle panel) Dependence of sSFR
on $M_\mathrm{HI}/M_\mathrm{*}$ for galaxies with strong bars, weak bars and unbarred galaxies.
(Bottom panel) Dependence of the bar fraction on t$_\mathrm{cons}$.
}\label{tcons}
\end{figure} 
 
In Figure \ref{fgas} top panel, we observe an anticorrelation between the bar fraction and the
HI gas mass fraction, particularly clear for strong bars, the fraction of weak bars
actually shows a positive correlation, with a mild increase of $f_\mathrm{bar}$ with
increasing  $M_\mathrm{HI}/M_\mathrm{*}$.
With the gas mass fraction decreasing for increasing stellar mass, the decrease of
$f_\mathrm{bar}$ with increasing $M_\mathrm{HI}/M_\mathrm{*}$ could be
a direct consequence of the dependence of $f_\mathrm{bar}$ on the stellar mass.
To test if $f_\mathrm{bar}$ depends directly on the HI gas mass fraction, we present
the bar fraction for strong bars in the $M_\mathrm{HI}/M_\mathrm{*}$ v. M$_\mathrm{*}$
plane in Figure \ref{fgas} bottom panel. The contours show a clear joint dependence
of $f_\mathrm{bar}$ on both, the stellar mass and the HI gas mass fraction, so even at
fixed M$_\mathrm{*}$, the strong bar fraction presents a clear dependence on
$M_\mathrm{HI}/M_\mathrm{*}$, as previously pointed out by Masters et al. (2012).
We are not including molecular gas in the discussion because the typical
molecular-to-atomic ratio is only $\sim$ 0.3, and given that the position of galaxies
when plotted in the SFR vs. M$_\mathrm{*}$ plane can be explained by their
global cold gas reservoirs as determined through the HI line \citep{Saintonge16},
we do not expect its inclusion to change our general conclusions.

Two frequently invoked explanations for this anticorrelation of the strong bar fraction with
HI gas mass ratio are, (i) that bars promote the consumption of atomic gas and, (ii) that
gas in disk galaxies inhibits the formation of bars and/or prevents their growth. Here,
we want to explore if any of these two hypothesis are able to explain the decrease of
the bar fraction with increasing HI gas mass fraction found in our sample.

If galaxies with strong bars consume faster their HI gas than unbarred ones, we expect
that for a given $M_\mathrm{HI}/M_\mathrm{*}$ ratio, galaxies with strong bars would
be consuming their gas at a higher rate than unbarred galaxies, presenting a higher
star formation activity. In Figure \ref{tcons} top panel, we plot the H$\alpha$ equivalent
width as a function of $M_\mathrm{HI}/M_\mathrm{*}$ for galaxies hosting strong bars,
weak bars and unbarred galaxies. We use $W(H\alpha)$ as an indicator of star formation
rate in the central region of galaxies ($R < 1.''5$). Within error bars, we do not find any statistical difference
for the value of $W(H\alpha)$ at fixed $M_\mathrm{HI}/M_\mathrm{*}$ between barred
and unbarred galaxies.

Figure \ref{tcons} middle panel shows the specific star formation rate (sSFR) as a function
of HI gas mass ratio for the three subsamples. We find that
the global sSFR increases for increasing $M_\mathrm{HI}/M_\mathrm{*}$, with unbarred
galaxies having systematically higher sSFR than strongly barred galaxies for a given
$M_\mathrm{HI}/M_\mathrm{*}$ value, a trend in the opposite direction of the one
expected if strong bars in galaxies promote gas consumption.

An alternative possibility to explore this hypothesis is looking at the bar fraction as a function of
the gas consumption timescale, defined as $t_\mathrm{cons} = M_\mathrm{HI}/SFR$
(Roberts 1963), which is the time a galaxy would take to consume its HI gas mass
if its star formation continues at the same rate as at present. It is important to keep in mind
that this is just a rough estimate of the time that galaxies can sustain certain amount of star formation.
Even in a close box scenario, given the tight correlation between the star formation rate and
the gas column density (Kennicutt 1998), a decrease in the gas column density due to star formation
would decrease the star formation which in turn decreases $t_\mathrm{cons}$. In a more realistic
scenario, gas can be accreated into galaxies, but unless barred and unbarred galaxies are accreating
gas at different rates, we still can compare gas consumption timescales for each sub-sample to draw 
general conclusions. If the trend between
$f_\mathrm{bar}$ and $M_\mathrm{HI}/M_\mathrm{*}$ is caused by the quickly consumption
of atomic gas in barred galaxies, the bar fraction should be higher in galaxies with
short gas consumption timescales. As we observe in Figure \ref{tcons} bottom panel, this is
not the case, the bar fraction presents only a weak dependence on t$_\mathrm{cons}$,
with a mild increase of $f_\mathrm{bar}$ with increasing t$_\mathrm{cons}$ for strong bars,
a trend in the opposite direction of the one expected according our hypothesis.

This quenching of star formation in barred galaxies is not unheard-of, and has actually
been recently reported by several authors (Wang et al. 2012; Cheung et al. 2013; Gavazzi et al. 2015),
specially for the case of massive galaxies, where the bar might have funneled a large proportion of gas
to the central region, causing a brief but strong burst of star formation, starving the outer regions and
turning the galaxy into a quiescent one.

The results presented in Figure \ref{tcons} favors the explanation where the low bar fraction in
HI gas rich galaxies is not a result of strong bars promoting the quick consumption of gas, but
as a result of the effect of gas on bar formation. The effect of the gas on bar formation and
evolution has been addressed by several theoretical studies. Berentzen et al. (2007), studying
the evolution of a live disk-halo system found that in gas-rich disks, the bar funnels gas to
the central region where it forms a central mass concentration (CMC), that in turn weakens the bar.
In the simulations by Villa-Vargas, Shlosman \& Heller (2010), bars are prevented to grow in
gas rich disks.
Athanassoula, Machado \& Rodionov (2013) reported that in their N-body simulations, long-scale
bars form later in gas-rich systems than in gas-poor ones, and that these bars are weaker in
gas-rich cases. They attribute this to the formation of CMCs that are formed in gas-rich systems
when gas is pushed inwards, also arguing that gas might transfer angular momentum to the
bar, hindering its growth.

Finally, as stars are formed in thin disks of gas that gradually thickens through dynamical effects,
gas rich galaxies are expected to present more prominent thin disks than gas poor galaxies.
Klypin et al. (2009) reported slowly rotating long bars in their thick disk models, in contrast with
the rapidly rotating shorter bars present in the thin disks, which following the previous argument
favors the formation of strong, slowly rotating bars in gas poor systems.

\section{Conclusions}

Using a sample of galaxies from the SDSS where bars are visually identified, and using 
HI mass and kinematic information from ALFALFA, we studied the fraction of galaxies
hosting bars as a function of stellar and baryonic masses, as well as two estimates
of halo mass. We found an increase of $f_\mathrm{bar}$ for strong bars, with increasing stellar and
baryonic masses, with a stronger dependence on M$_\mathrm{*}$ than on M$_\mathrm{baryonic}$.
The signal of the correlation of the bar fraction with the halo mass estimates is weak.

We confirm previous results by CS+15 and D\'iaz-Garc\'ia et al.
(2016), finding an increase of the strong bar fraction
with increasing stellar-to-halo mass ratio. For massive galaxies in our sample, with $M_\mathrm{*} >
10^{10} M_{\odot}$, the dependence
of the bar fraction on M$_\mathrm{dyn}$ and M$_\mathrm{halo}$
is present even at fixed M$_\mathrm{*}$, with decreasing $f_\mathrm{bar}$ for increasing
global halo mass, measured to a distance of the order of the HI disk extent. Compared
with the dependence of $f_\mathrm{bar}$ on stellar mass, the dependence of $f_\mathrm{bar}$ on the 
dynamical and halo masses is small. 

We find a strong correlation between the bar fraction and the HI gas mass ratio,
such that the strong bar fraction decreases with increasing atomic gas content, in good
agreement with Masters et al. (2012). The dependence of $f_{\mathrm{bar}}$
with $M_\mathrm{HI}/M_\mathrm{*}$ is usually explained invoking two mechanisms:
(i) strong bars promote the consumption of atomic gas and, (ii) gas prevents
the formation/growth of bars. Our results show that barred galaxies in our
sample are not consuming their gas in a more efficient way than their unbarred
counterparts, hence favoring the second explanation; increasing the gas content
in disk galaxies prevents the formation of bars, they grow more slowly or
they are destroyed directly or indirectly by the presence of gas,
as explained by recent theoretical works (Berentzen et al. 2007; Villa-Vargas,
Shlosman \& Heller 2010;
Athanassoula, Machado \& Rodionov 2013;  Algorry et al. 2016).

\acknowledgments
The author thanks Changbom Park for providing the sample for the present study, and
S. Courteau and IRyA Extragalactic group for valuable discussions about the results.
The author also thanks the anonymous referee for useful comments 
that helped to improve the quality of the paper and clarify the results.
    Funding for the SDSS and SDSS-II has been provided by the Alfred P. Sloan Foundation,
    the Participating Institutions, the National Science Foundation, the U.S. Department of
    Energy, the National Aeronautics and Space Administration, the Japanese
    Monbukagakusho, the Max Planck Society, and the Higher Education Funding Council
    for England. The SDSS Web Site is http://www.sdss.org/. The SDSS is managed by the
    Astrophysical Research Consortium for the Participating Institutions. The Participating
    Institutions are the American Museum of Natural History, Astrophysical Institute Potsdam,
    University of Basel, University of Cambridge, Case Western Reserve University,
    University of Chicago, Drexel University, Fermilab, the Institute for Advanced Study, the
    Japan Participation Group, Johns Hopkins University, the Joint Institute for Nuclear Astrophysics,
    the Kavli Institute for Particle Astrophysics and Cosmology, the Korean Scientist Group,
    the Chinese Academy of Sciences (LAMOST), Los Alamos National Laboratory,
    the Max-Planck-Institute for Astronomy (MPIA), the Max-Planck-Institute for Astrophysics (MPA),
    New Mexico State University, Ohio State University, University of Pittsburgh,
    University of Portsmouth, Princeton University, the United States Naval Observatory,
    and the University of Washington.

\end{document}